\documentclass{singlecol-new}
\usepackage[numbers, square, comma, sort&compress]{natbib}
\usepackage{natbib,stfloats}
\usepackage{subfigure}
\usepackage{mathrsfs}
\usepackage{tabu}
\usepackage{graphicx}
\def\newblock{\hskip .11em plus .33em minus .07em}

\theoremstyle{TH}{

}

\theoremstyle{THrm}{

}

\theoremstyle{THhit}{

}

\makeatletter
\def\theequation{\arabic{equation}}

\makeatother

\usepackage{url}
\usepackage{graphics}
\usepackage{graphicx}
\begin{document}%

\setcounter{page}{1}

\LRH{S. Shehab et~al.}

\RRH{A Survey of the State-of-the-Art Parallel Multiple Sequence Alignment Algorithms on Multicore Systems}

\VOL{x}

\ISSUE{x}

\PUBYEAR{201X}

\BottomCatch

\CLline

\subtitle{}

\title{A Survey of the State-of-the-Art Parallel Multiple Sequence Alignment Algorithms on Multicore Systems}


%
\authorA{Sara Shehab}
\affA{Computer Science Department,\\ Faculty of Computers and Information,\\Menoufia University, Shebeen El-Kom, 32511, Egypt \\
E-mail: sara.Shehab@ci.menofia.edu.eg}
\authorB{Sameh Abdulah}
\affB{Computer Science Department,\\ Faculty of Computers and Information,\\Menoufia University, Shebeen El-Kom, 32511, Egypt \\
E-mail: sameh.abdulah@ci.menofia.edu.eg}
\authorC{Arabi E. Keshk}
\affC{Computer Science Department,\\ Faculty of Computers and Information,\\Menoufia University, Shebeen El-Kom, 32511, Egypt \\
E-mail: arabi.keshk@ci.menofia.edu.eg}

%
%
%
%
%
%
%

\begin{abstract}
Evolutionary modeling applications are the best way to provide full
information to support in-depth understanding of evaluation of organisms.
These applications mainly depend on identifying the evolutionary history
of existing organisms and understanding the relations between them,
which is possible through the deep analysis of their biological sequences. 
Multiple Sequence Alignment (MSA) is considered an important tool in such
applications, where it gives an accurate representation of the
relations between different biological sequences. In literature, many efforts have been
 put into presenting a new MSA algorithm
or even improving existing ones. However, little efforts on optimizing parallel
MSA algorithms have been done. Nowadays, large datasets become a reality, 
and big data become a primary challenge in various fields, which should be
also a new milestone for new bioinformatics algorithms.

This survey presents four of the state-of-the-art parallel MSA algorithms,  
{\em T-Coffee}, {\em MAFFT}, {\em MSAProbs}, and {\em M2Align}. We provide a detailed discussion of 
each algorithm including  its strengths, weaknesses, and implementation details
 and the effectiveness of its parallel implementation compared to the other algorithms,
 taking into account the  MSA accuracy on two different datasets,  BAliBASE and
 OXBench. 

\end{abstract}

\KEYWORD{Bioinformatics; Multiple Sequence Alignment; Parallel Processing; Multicore Systems.}


\begin{bio}
\textbf{Sara Shehab} is  a Ph.D. student at computer
science department, faculty of computers and information,
Menoufia University. Sara received her
MS from Faculty of computers and information,
Menoufia university, Egypt, 2012. Mainly, 
her work centered around pair-wise and multiple
sequence alignment in bioinformatics.

\noindent \textbf{Sameh Abdulah} is an assistant professor at
computer science department, faculty of computers
and information, Menoufia University. Sameh received
his MS and PhD degree from The Ohio State
University, Columbus, USA, 2012 and 2016, his
work is centered around High Performance Computing
(HPC) Machine Learning and Data Mining
(MLDM) applications algorithms., and bioinformatics
applications.

\noindent \textbf{Arabi E. keshk} is a professor of computer
science at faculty of computers and information,
Menoufia University, he received the B.Sc. in Electronic
Engineering and M.Sc. in Computer Science
and Engineering from Menoufia University, Faculty
of Electronic Engineering in 1987 and 1995, respectively
and received his PhD in Electronic Engineering
from Osaka University, Japan in 2001. His
research interest includes software testing, digital
systems testing, cloud computing, data science, and
bioinformatics.
\end{bio}

\maketitle

\section{Introduction}
\label{sec:intro}
For several years, studying the evolution of organisms plays an important role to understand their life, development, and physical structure. This kind of study requires an in-depth analysis of existing biological data to gain more information about different organisms. Thus, bioinformatics science is arisen as a new computation field to define a set of algorithms, methods, and techniques for understanding biological data~\cite{zomaya2006parallel}. Biological data is usually massive and requires developing and applying computationally intensive techniques. 
 
 One of the most challenging problem in bioinformatics is to extract the evolutionary relationship between different organisms. Assuming a set of biological sequences protein, DNA, or RNA sequences, the relation between two or more sequences can be shown as a sequence alignment process. Such a process is a fundamental tool in several applications, molecular function prediction, intermolecular interactions, residue selection, and phylogenetic analysis.

In this study, we concentrate on the Multiple Sequence Alignment (MSA) problem. In literature, different sequential algorithms have been proposed to solve such a problem based on different methods. For example, progressive methods (ex., T-Coffee~\cite{notredame2000t}, Clustal), Multiple Alignment using Fast Fourier Transform (MAFFT)~\cite{katoh2013mafft}, Iterative methods (ex., MUSCLE~\cite{edgar2004muscle}), Consensus methods (ex., M-COFFEE~\cite{wallace2006m}),  Hidden Markov models (i.e., HMMER~\cite{finn2011hmmer}), an the intuitive methods (ex. PoMSA~\cite{shehab2017pomsa}). With the rapid growth of sequence databases, which now contains enough representatives of larger protein families to exceed the capacity of most current programs, the complexity of existing sequential algorithms is increased even in the case of two sequences alignment. For example, computations of current homologous sequence datasets could take several days.

In fact, the best methods sometimes fail to deal with these complexities efficiently and obtain biologically accurate alignments at the same time. The present studies overcome these obstacles by using two main approaches. The first is the vectorization, where all matrices are compensated by vectors, which in turn reduces the memory requirement and speed up execution without affecting the accuracy. The second approach is parallelism, the widespread programming method nowadays that allows multiple independent processes which share the same resources, to be executed concurrently at less time. 

This paper concentrates on the parallel approach by studying four different shared-memory parallel implementations of the current state-of-the-art MSA algorithms, i.e., {\em T-Coffee}~\cite{notredame2000t, di2010cloud}, {\em MAFFT}~\cite{katoh2013mafft, katoh2010parallelization}, {\em MSAProbs}~\cite{liu2010msaprobs}, and {\em M2Align}\cite{zambrano2017m2align} to show the strengths and weaknesses of each implementation. The importance of such a study is to illuminate the road for researchers to improve the existing implementation of these algorithms and provide new parallel solutions which became necessary with the sizable biological data we have today.

\section{Parallel Alignment Algorithms Overview}
\label{sec:palgorithm}
Although dynamic programming is the most optimal MSA solution according to
the accuracy levels it can obtain, alignment several numbers of sequences
is prohibitive because of high computation requirement of such a 
solution~\cite{notredame2002recent, liu2010msaprobs, shehab2017pomsa}.
Therefore, in literature, many heuristics  provide a more practical solution
with less computation complexity, such as progressive alignment~\cite{feng1987progressive},
iterative alighnment~\cite{barton1987strategy}, and  alignment based on the profile
Hidden Markov Models (HMM)~\cite{eddy1998profile}.  Parallel algorithms also suffer from
the memory bound restrictions that are provided by each algorithm. Thus, existing
parallel implementations on shared memory mostly use MSA progressive alignment
strategy to avoid memory overflow problems.

Here, we  give a deep overview of four different parallel algorithms,
{\em T-Coffee}, 
{\em MAFFT}, 
{\em MSAProbs}, and {\em M2Align}.
The overview  will cover both the sequential and parallel implementations for each algorithm.

\subsection{T-Coffee Algorithm}
Tree-based Consistency Objective Function for Alignment Evaluation
algorithm ({\em T-Coffee}), is a progressive MSA algorithm, which optimizes
the original progressive alignment in {\em Clustalw} algorithm~\cite{thompson2002multiple}.
The optimization involves using pre-aligned pair-wise sequences output (i.e., {\em T-Coffee}
initial sources) from two or more pre-selected algorithms to improve the overall alignment
operation. Even in the case of inconsistent pair-wise alignments in different sources,
{\em T-Coffee} is able to differentiate the alignments of the same residue pairs by associating
them with different weights.

The earliest implementation of {\em T-Coffee} algorithm has been proposed
by~\cite{notredame2000t, thompson2002multiple}. The proposed implementation
has used two different pairwise alignment algorithms to initially aligned the
given sequences, {\em ClustalW}, i.e., global pairwise aligner, and
{\em lalign}, i.e., local pairwise aligner. Both algorithms generate a separate
source of pairwise aligned sequences.  {\em T-Coffee} assigns a separate weight to each
aligned residue pair, which represents the correctness of the alignment before
merging different sources as one primary library. During the merging step, {\em T-Coffee}
can find the same residue pair with different alignment coming from different sources,
in this case, each distinguish alignment should be represented as a separate
entry in the {\em T-Coffee} primary library. However, if the same residue pair appears more
than once in different sources with the same alignments, it should be added once
to the primary library with a weight equals to the sum of the two weights.

The {\em T-Coffee} primary library can be used directly to perform MSA for the
given sequences, however, more optimization can improve the whole alignment operation.
For instance, another extension step can be applied to the primary library by adding
another weight score that shows how a residue pair align is consistent with other sequences.
In this case, the most consistent residue alignments are selected to be used in the MSA alignment
step. The extended library now contains all the information that can improve the overall MSA
alignment operation by using a traditional progressive alignment process.

In the progressive alignment technique, pairwise alignment library should be built first. 
In the case of the {\em T-Coffee} algorithm, the extension library can be used directly. 
{\em T-Coffee} depends on the dynamic programming to apply the progressive
alignment strategy. The {\em T-Coffee} algorithm generates a distance matrix based
on the extension library. The distance matrix is used to generate the guide tree
that helps in generating the final list of aligned sequences.


In this study, we concentrate on the parallel implementations. 
In~\cite{di2010cloud}, a parallel implementation of the {\em T-Coffee}
algorithm has been presented. The proposed parallel implementation
targets four main parts of the algorithm, template selection, library computation, 
library extension, and progressive alignment. Template selection is a new feature
added to most advanced modes of T-Coffee (i.e., R-Coffee, 
3D-Coffee~\cite{armougom2006expresso}, and PSI-Coffee~\cite{chang2012accurate})
where input sequences are associated with structural templates~\cite{di2010cloud}.
In this study, we concentrate on the last three parts which are shown in the
{\em T-Coffee} base algorithm.

The library computation step is related to the use of different pairwise alignment algorithms
to provide a set of initial libraries to the  {\em T-Coffee} algorithm.  This step can be parallelized
by dividing the alignment task into several tasks that are equal to the number of available cores.
The master process merges the outputs coming from different cores into a single library. 
 
\begin{figure}
\centering
\includegraphics[width=\linewidth]{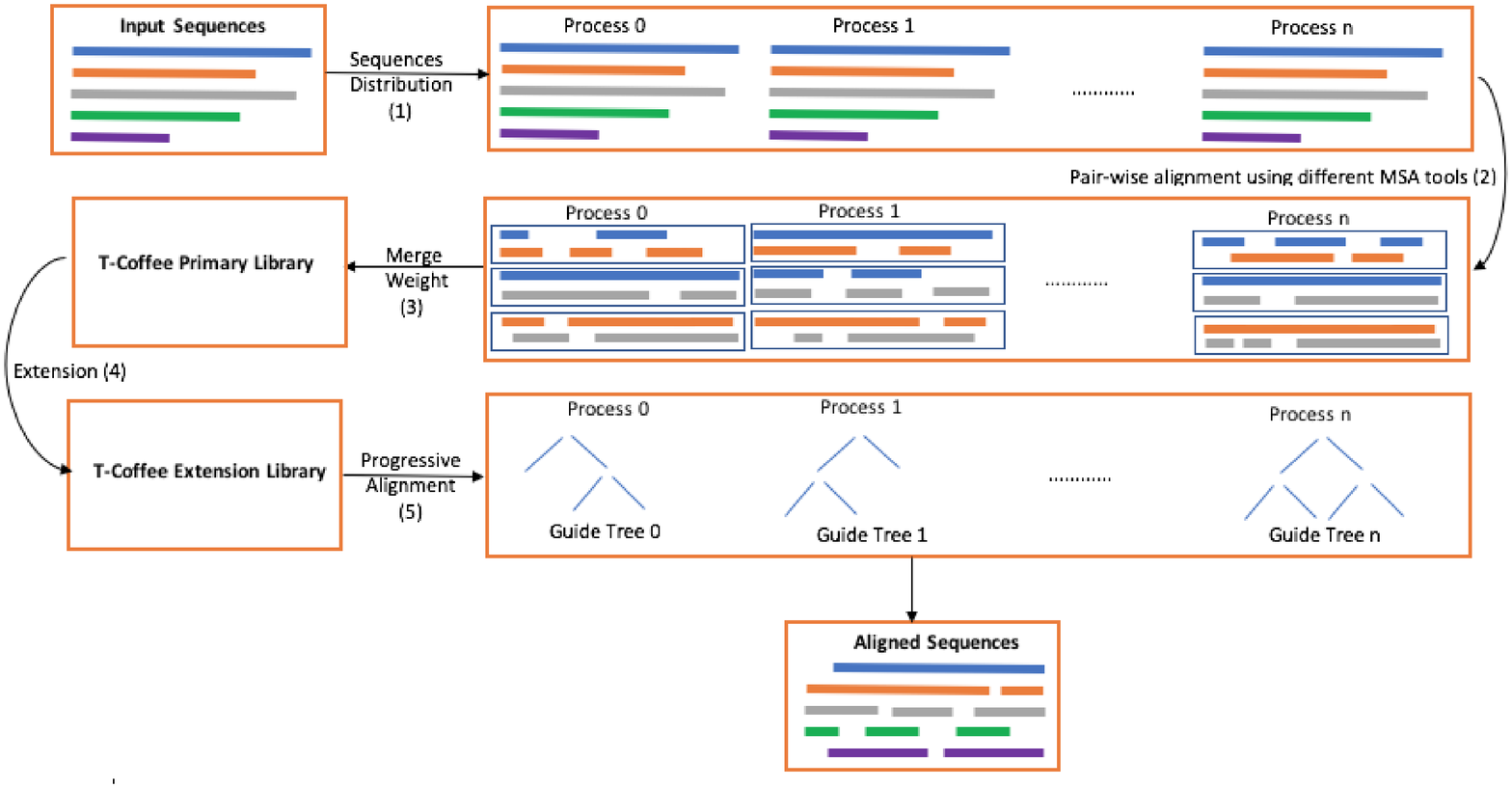}
  \caption{Parallel T-coffee algorithm.}
  \label{fig:tcoffee-fig}
\end{figure}

The extension library generation step can also be parallelized by distributing the residue pairs
to the available cores so that extended weights can be added simultaneously.  Moreover, the
progressive alignment step is parallelized by processing the guide tree with different processes.
Each process deals with only one node independently to generate the final output. 
Figure~\ref{fig:tcoffee-fig} gives an overview of parallel implementation of {\em T-Coffee} algorithm.

\subsection{MAFFT Algorithm}

 \begin{figure}
\centering
\includegraphics[width=0.90\linewidth]{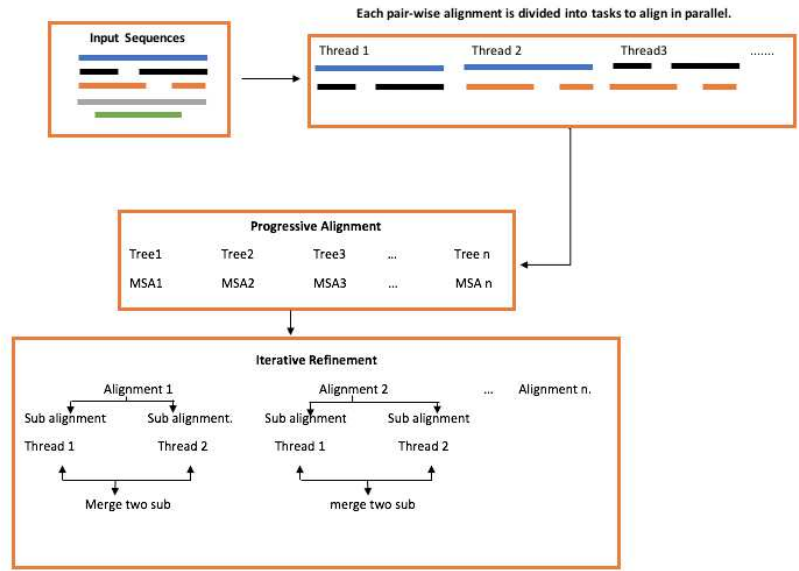}
  \caption{Parallel MAFFT algorithm.}
  \label{fig:mafft-fig}
\end{figure}

Multiple  sequence Alignment based on Fast Fourier Transform ({\em MAFFT}), is an 
MSA algorithm which depends on the fast Fourier transform algorithm to fast discover
matched parts of a given set of sequences. {\em MAFFT} has three heuristic running
modes, progressive method (FFT-NS-2), the iterative refinement method
(FFT-NS-i, L-INS-i, E-INS-i, and G-INS-i), and  structural alignment method for 
RNA (Q-INS and X-INS-i)~\cite{katoh2013mafft}. Each mode is suitable for a certain
number of sequences and the user priority (i.e., fast or accurate computation).

The progressive method in {\em MAFFT} is a straightforward method of generating
the distance matrix and the guide tree to better align the given set of sequences.
This option usually used in the case of a very large number of sequences and fast
computation priority from the user. In the iterative refinement method, accurate alignment
is considered a priority and several alignment iterations are required. Further, structural
alignment method is more suitable for RNA alignment proteins with low sequence similarity.

Sequential {\em MAFFT}  has three main steps, all-to-all comparison, progressive alignment,
and iterative refinement.  The all-to-all comparison can be considered as a part of the progressive
alignment step where a set of pairwise comparisons are conducted to generate the initial distance
matrix and the guide tree that are used by the progressive alignment step. Progressive alignment
is the main MSA process while iterative refinement step is used to improve the accuracy of the final result.

In~\cite{katoh2010parallelization} a parallel {\em MAFFT}  algorithm has been proposed.
The given implementation is based on POSIX Threads library and targets the three main steps
of the {\em MAFFT} algorithm. All-to-all comparison has been parallelized by distributing the pairwise
alignment to the available physical cores. In progressive alignment, parsing the guide tree is
parallelized every level, i.e., each tree parent depends on its child which prevents parallelization
over the whole tree nodes. Finally, if iterative refinement is used, two different approaches have
been proposed,  {\em best-first approach} and  {\em hill-climbing approach}, where both of them
aim at dividing each alignment into two sub-alignments and the two sub-alignments are realigned.
The realignment step is performed  according to the tree dependent iterative strategy proposed
in~\cite{gotoh1993optimal}.

In the {\em best-first approach}, The realignments are performed for all possible alignments on the
iterative tree. The alignment with the highest objective score is selected for another iteration. Once
no high score alignment is found, the algorithm terminates. In the {\em hill-climbing} approach, each
process has its local realignments of the original alignment and the better score is replaced the original
alignment is kept locally by each process. In this case, many realignments of the same alignment are
used in next iteration~\cite{katoh2010parallelization}.  Figure~\ref{fig:tcoffee-fig} gives an overview of
parallel { \em MAFFT} algorithm.

\subsection{MSAProbs}
 \begin{figure}
\centering
\includegraphics[width=0.90\linewidth]{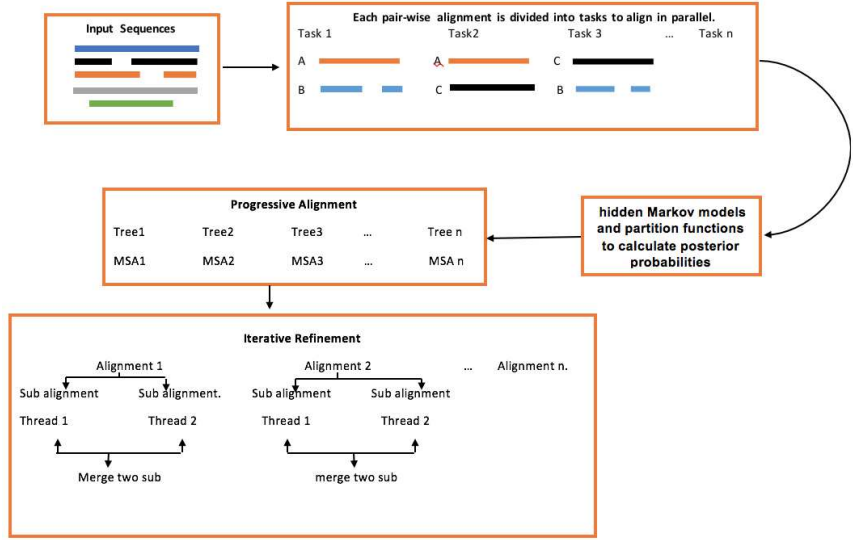}
  \caption{Parallel MSA-Probs algorithm.}
  \label{fig:msaprobs-fig}
\end{figure}

MSAProbs  is a progressive algorithm which provides an effective solution
to perform MSA by combining both Hidden Markov Models (HMM) technique
with partition functions to calculate posterior alignment probabilities. MSAProbs
is able to provide a higher alignment accuracy through two new techniques,
weighted probabilistic consistency transformation and weighted profile-profile
alignment~\cite{liu2010msaprobs}.

Sequential MSAProbs consists of five main stages, generate pairwise posterior 
probability matrices using  a pair-HMM and a partition function, generate a pairwise
distance matrix using the set of posterior probability matrices, constructing a  guide tree
from the pairwise distance matrix to estimate different sequences weights, performing
a weighted probabilistic consistency transformation of all pairwise posterior probability
matrices, and apply a progressive alignment on the {\em guide tree} using the transformed
posterior probability matrices. In-depth details about each step is shown in~\cite{liu2010msaprobs}.

In this paper, we concentrate more on the parallel implementation of MSAProbs which
is available under MSAProbs v0.9.7 package~\cite{msaprobswebpage}. Among the five
MSAProbs' steps, generating the pairwise posterior probability matrices, i.e., step 2, 
 and weighted probabilistic consistency transformation, i.e., step 4 , are the most time
consuming part with time complexity
of $\mathcal{O}(N^2L^2)$ and $\mathcal{O}(N^2L^3)$, respectively. $N$ represents the number of
sequences and $L$ represents the average length of sequences.

Current implementation of parallel MSAProbs is mostly bases on parallizing these two most
consuming steps in the algorithm using OpenMP, i.e., a compiler-directive-based API for 
shared-memory parallelism~\cite{dagum1998openmp}. Figure~\ref{fig:msaprobs-fig} shows an
 overview of parallel MSA-Probs algorithm.

\subsection{M2Align}
Multi-objective metaheuristics methods is another approach to deal with MSA problems.
In such methods, a non-exact stochastic optimization function is used to highly optimized
two or more objectives concurrently. A Matlab-based tool called MO-SAStrE is an example
of applying such an approach to the MSA problem~\cite{ortuno2013optimizing}.
This tool depends on genetic algorithms through a set of mutation and crossover operations
to the expected alignment results to find the best alignment solution for a set of given sequences.
NSGA-II genetic algorithm is the backbone of MO-SAStrE tool~\cite{deb2002fast}.

The {\em M2align} algorithm is a parallel multiple sequence alignment algorithm based on
MO-SAStrE.  Besides exploiting the multi-core architecture of current computing devices
to parallelize the MO-SAStrE algorithm, it optimizes the functionality of the algorithm by
using better storage mechanism to store gaps information during the optimization phase.

\begin{figure}
\centering
\includegraphics[width=0.90\linewidth]{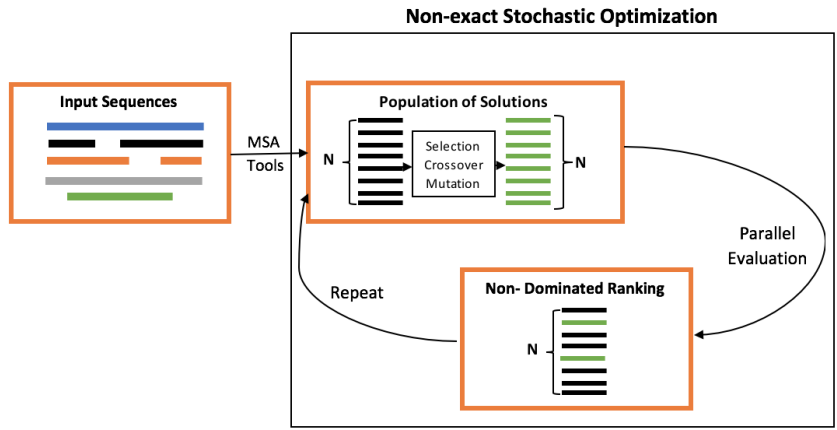}
  \caption{Parallel M2align algorithm.}
  \label{fig:m2align-fig}
\end{figure}

The {\em M2align} algorithm is working as follows. A set of pre-determined MSA algorithms
is used to generate $N$ alignment solutions which are used as the initial population
in the optimization phase. Another set of solutions can be added to the population
through applying genetic crossover operators (i.e., Single-Point Crossover) to each
pair of the initial N solutions. later, a set of $N$ solutions is selected based on its
dominance ranking~\cite{zambrano2017m2align}. An initial evaluation step is applied to
assign a dominance rank for each set of sequences. Figure~\ref{fig:m2align-fig} illustrates
the whole functionality of {\em M2align} algorithm over a given set of sequences.

The evaluation step in MO-SAStrE is a sequential step which takes a longer time to complete
compared to the other algorithm steps. Thus, M2align algorithm includes a parallel implementation
of this step to speedup the alignment process through multi-core systems.
In this case,  parallelization has been exploited by evaluating solutions
concurrently at the same time using differently available cores.

\section{Performance Evaluation}
\label{sec:perf}
In this section, we present a detailed performance evaluation
of the four parallel MSA algorithms: {\em T-Coffee}, {\em T-MAFFT},
{\em MSAProbs}, and {\em M2Align}. The evaluation involves execution
time and the obtained accuracy of these algorithms using a different number
of cores on two different datasets.

\subsection{setup}
\subsubsection{Experimental Testbed:}

We evaluate the performance of the target software packages on
a dual-socket 14-core Intel Broadwell Intel Xeon E5-2680 V4 processor
running at 2.4 GHz. The software packages are compiled with gcc v4.8
on Ubuntu 16.04.3 LTS. 

Two datasets have been used for evaluation, BAliBASE~\cite{thompson1999balibase},
and  OXBench~\cite{raghava2003oxbench}.  To validate the accuracy of different MSA
algorithms, we use {\em Bench}~\cite{bench},
which provides different benchmarks for several proteins datasets.

\subsection{Accuracy assessment}

We uses Three different metrics to evaluate the alignment accuracy of the given MSA algorithms,

\begin{enumerate}

\item {\em Q/TC score}:  $Q$ score represents the number of correctly aligned residue
pairs divided by the number of residue pairs in the reference alignment, i.e., Bench.
$TC$, i.e., total column score, represents the number of correctly aligned columns
divided by the number of columns in the reference alignment~\cite{edgar2004muscle}.
 
\item {\em The Cline's score (CS) or shift score}, this score is mainly represents a distance-based
scoring. It is negatively impacted  by both over-alignment where non-homologous regions are aligned
or  under-alignment where homologous regions are not aligned improbably~\cite{cline2002predicting}.
CS is considered as the best accuracy metric compared to existing scoring 
metrics~\cite{edgar2004comparison,lyras2014reformalign}.

\item {\em The modeler's score}  represents the number of correctly aligned  pairs in the aligned
file  divided by the number of pairs in the reference file~\cite{lyras2014reformalign}.

\end{enumerate}

\subsection{Execution time Performance Comparison}
Most of the existing MSA studies concentrate on
the alignment accuracy level that can be obtained by the given
MSA algorithm. However, in this study, we are mostly focusing
on the parallel execution time which should be associated with
an acceptable scalability trait. Scalability is considered an important
factor that shows the ability of a software to handle a growing amount
of work with a larger number of processes.

The following experiments show the performance of our target parallel MSA
algorithms using a different number of threads (i.e., 2, 4, 8, and 16) besides the
sequential implementation of each algorithm. We assume each core hosts only
one running thread (i.e.,  hyper-threading is disabled). Our target Broadwell 
multi-core system has a total of 28 cores. However, due to scalability issue of 
the existing software implementations, we perform our experiments with up to 16 threads.

\subsubsection{Parallel T-Coffee Algorithm Performance}

\begin{figure}[!ht]
  \centering
  \subfigure[Parallel T- Coffee - Balibase Dataset.]{
  \label{fig:tcoffe-balibase}
    \includegraphics[width=0.45\linewidth]{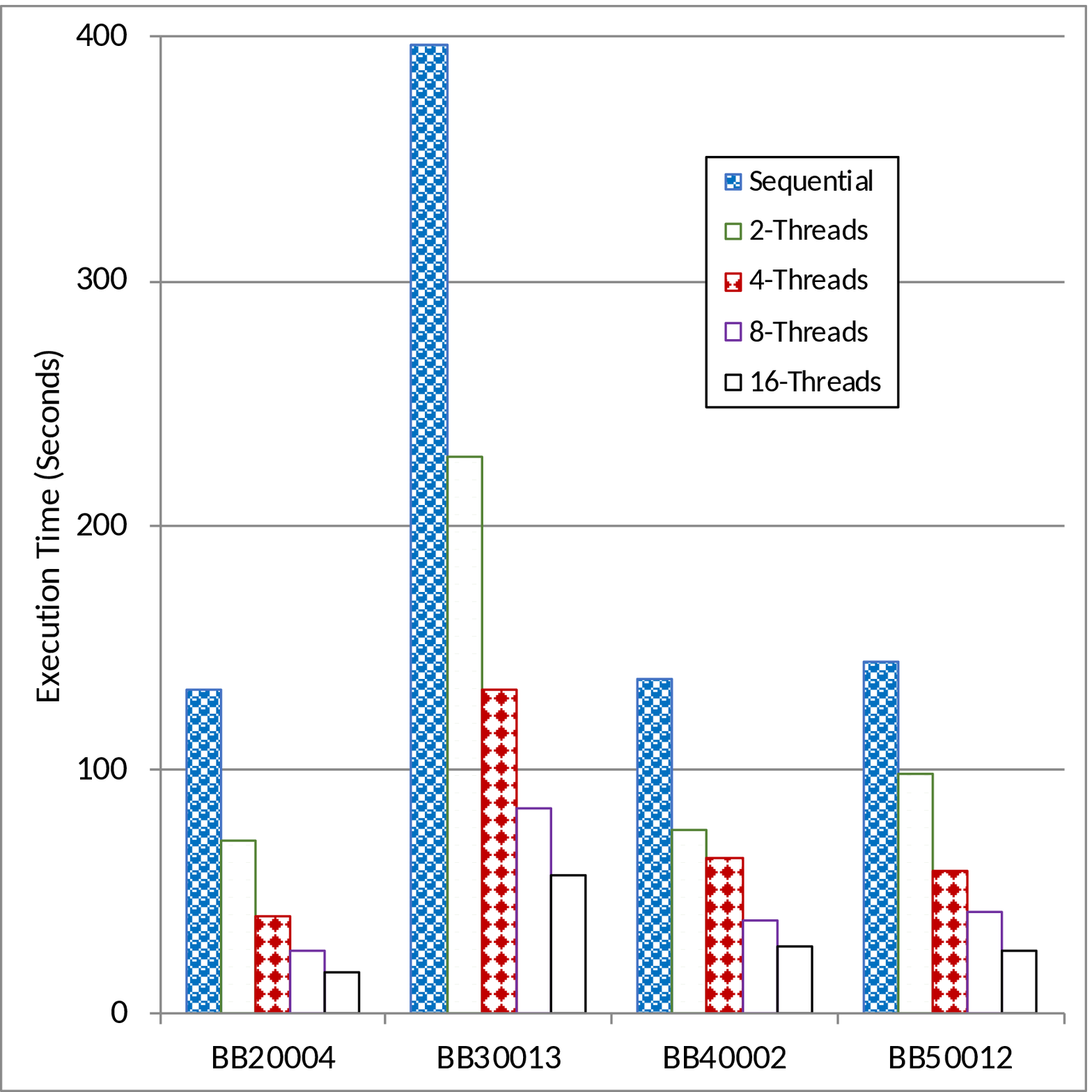}
    }
  \hfill
  \subfigure[Parallel T- Coffee - OXbench Dataset.]{
    \label{fig:tcoffe-oxbench}
   \includegraphics[width=0.45\linewidth]{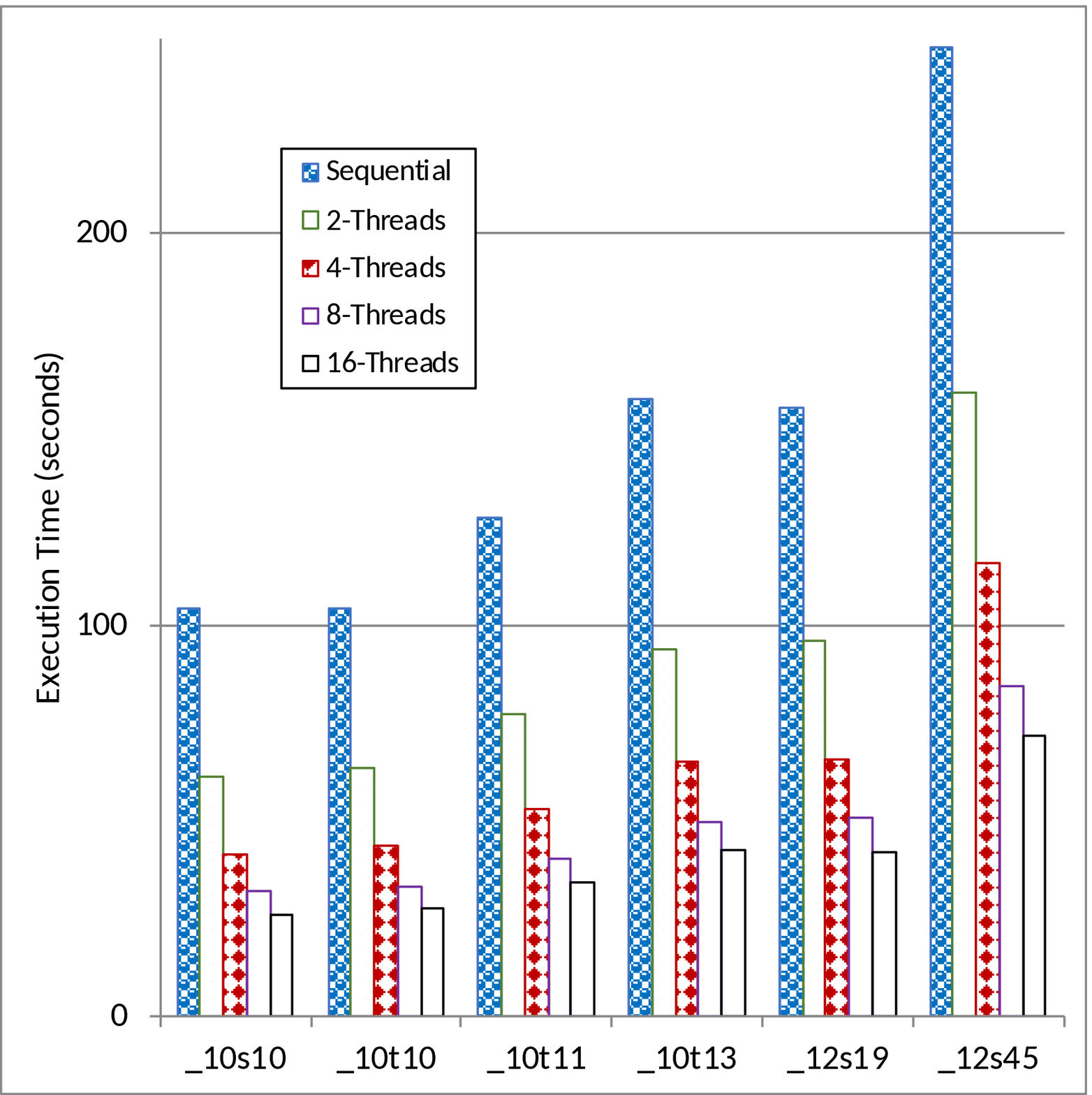}}

\caption{Parallel T- Coffee total execution time with different number of threads.}
\label{fig:execution-time}
\end{figure}

We evaluate {\em T-Coffee} implementation using 1, 2, 4, 8, and 16 threads.
Figures~\ref{fig:tcoffe-balibase} and~\ref{fig:tcoffe-oxbench}  show the
execution scalability using Balibase and OXbench datasets, respectively.
The figures shows linear scalability of T-Coffee with larger number of threads.
For example, in the case of BAliBASE dataset, the speedup from sequential execution
to 16 threads execution are $6.8X$, $6.1X$, $3.9X$, $4.5X$ in $BB20004$, $BB30013$,
$BB40002$, $BB50012$ files, respectively.	The same observation can be drawn from
Figure~\ref{fig:tcoffe-oxbench}. The speedup in the case of OXbench datasets
are $3X$, $2.8X$, $2.8X$, $2.7X$, $2.7X$, $2.4X$ in $\_10s10$, $\_10t10$, $\_10t11$,
$\_10t13$, $\_12s19$, $\_12s45$ files, respectively.

\subsubsection{Parallel MAFFT Algorithm Performance}
	
\begin{figure}[!ht]
  \centering
  \subfigure[Parallel MAFFT - Balibase dataset.]{
  \label{fig:tcoffee-balibase}
    \includegraphics[width=0.45\linewidth]{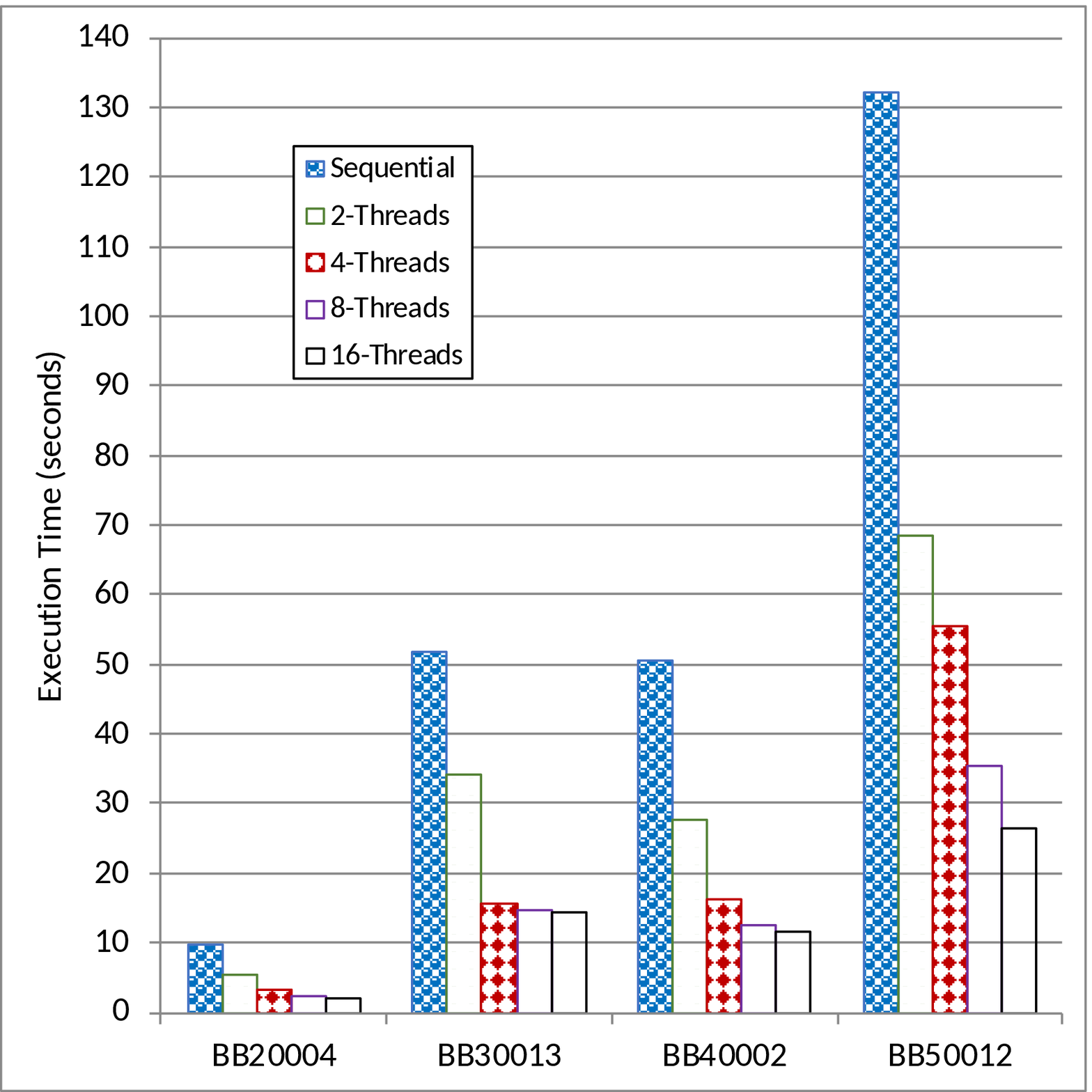}
    }
  \hfill
  \subfigure[Parallel MAFFT - OXbench dataset.]{
    \label{fig:tcoffee-oxbench}
   \includegraphics[width=0.45\linewidth]{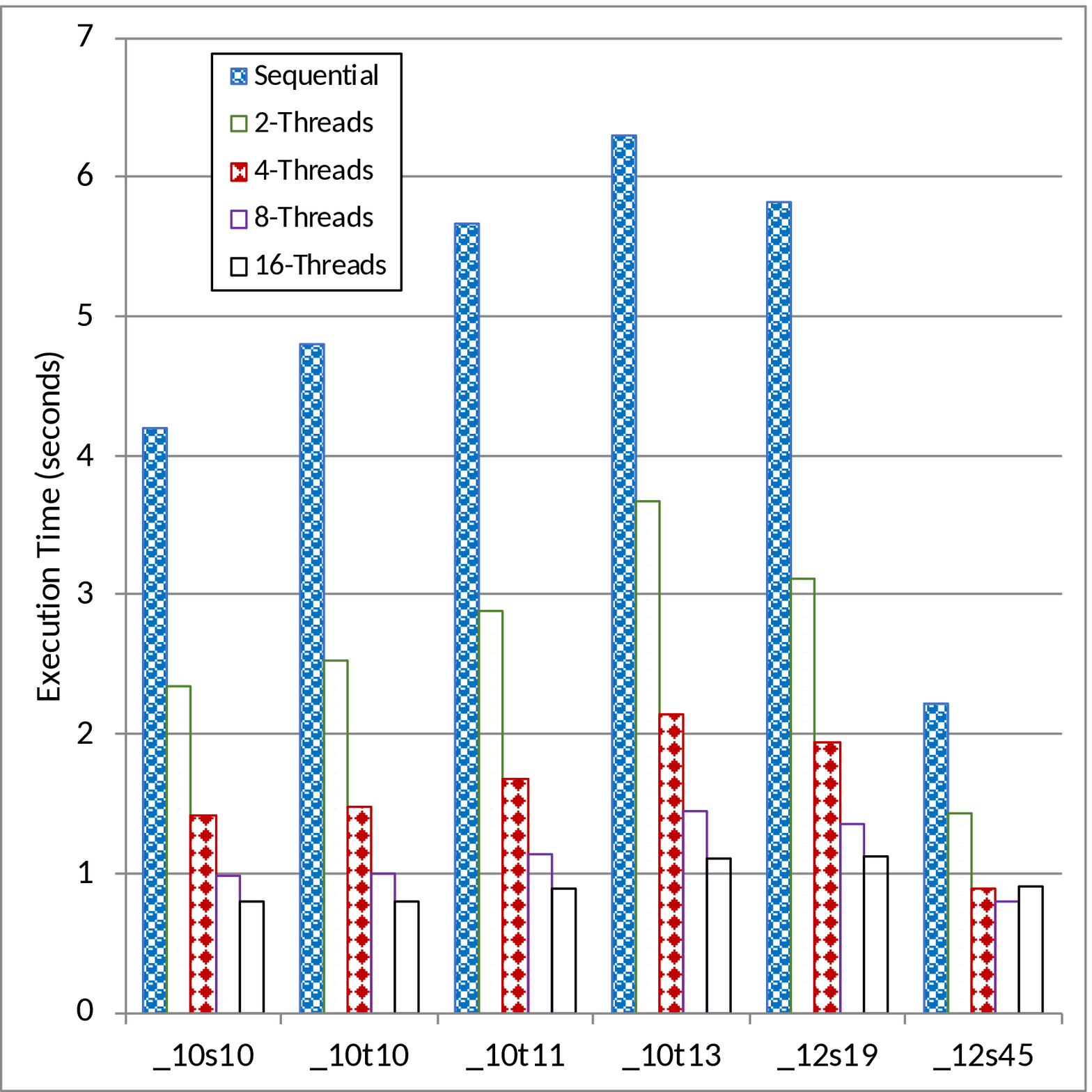}}

\caption{Parallel MAFFT total execution time with different number of Threads.}
\label{fig:execution-time-mafft}
\end{figure}

The same set of experiments have been conducted	using parallel {\em MAFFT}.
Figure~\ref{fig:execution-time-mafft} shows the execution time using both Balibase
and OXbench datasets.

{\em MAFFT} has three running options, i.e., progressive , iterative refinement,
and structural alignment. According to the dataset size only one option can be
chosen. Here, we choose an {\em auto} option, which allows the algorithm to
pick the best option for each dataset.

Figure~\ref{fig:tcoffee-balibase} shows the performance on Balibase dataset.
The gained speedup from sequential execution to 16 thread execution is
$3.73X$, $2.56X$, $3.32X$,and $4.02X$ in $BB20004$, $BB30013$, $BB40002$,
$BB50012$ files, respectively.  Figure~\ref{fig:tcoffee-oxbench} shows the execution
time using OXbench dataset. The speedup from sequential to 16 threads execution
is $4.22X$, $5.01X$, $5.34X$, $4.68X$, $4.17X$, and $1.44X$ in $\_10s10$, $\_10t10$, $\_10t11$,
$\_10t13$, $\_12s19$, $\_12s45$ files, respectively. The parallel {\em MAFFT} scalability
can be shown by both figures.

\subsubsection{Parallel MSAProbs Algorithm Performance}

\begin{figure}[!ht]
  \centering
  \subfigure[Balibase dataset with parallel MSAProbs.]{
  \label{fig:msaprobs-balibase}
    \includegraphics[width=0.45\linewidth]{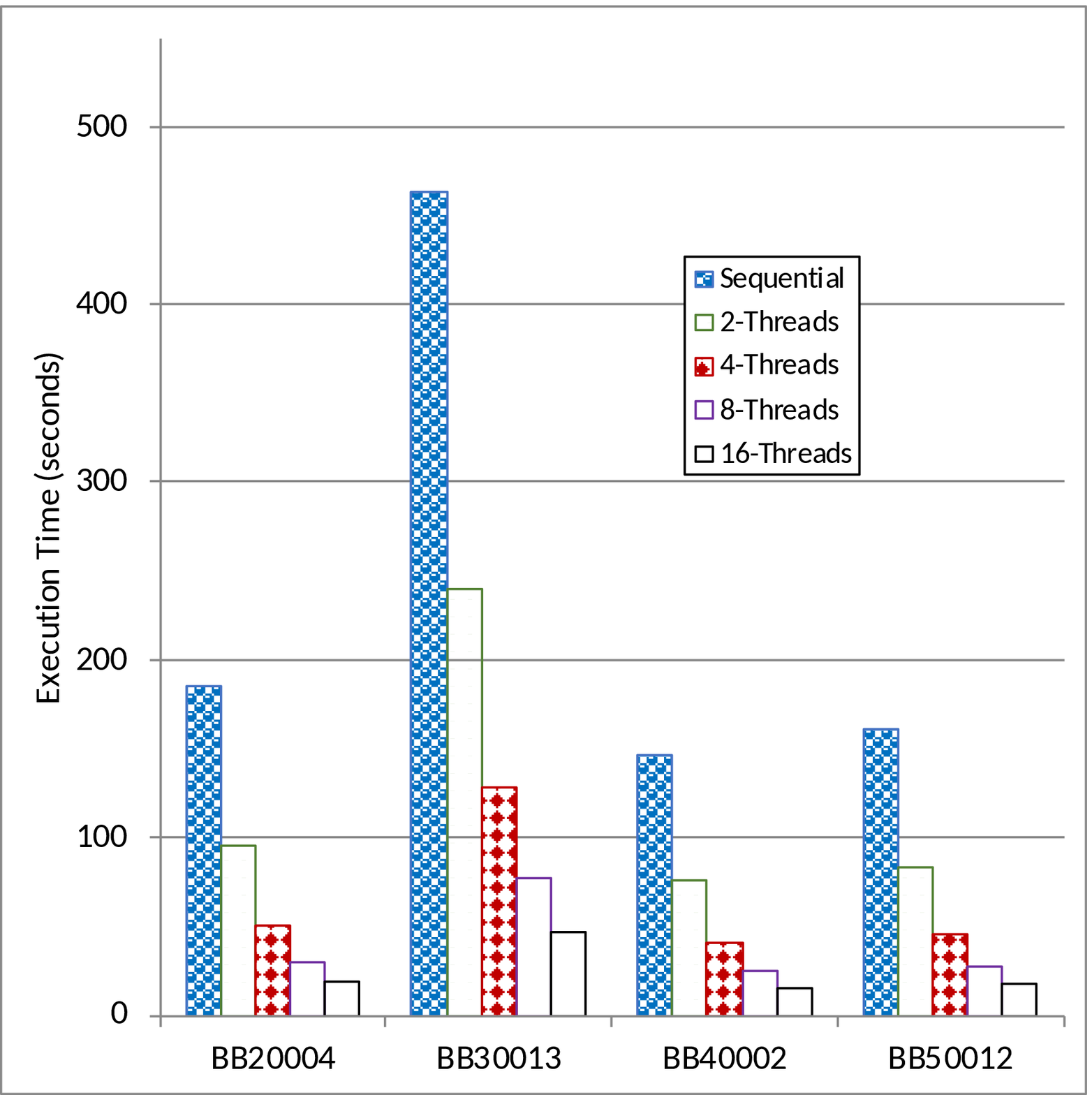}
    }
  \hfill
  \subfigure[OXbench dataset with parallel MSAProbs.]{
    \label{fig:msaprobs-oxbench}
   \includegraphics[width=0.45\linewidth]{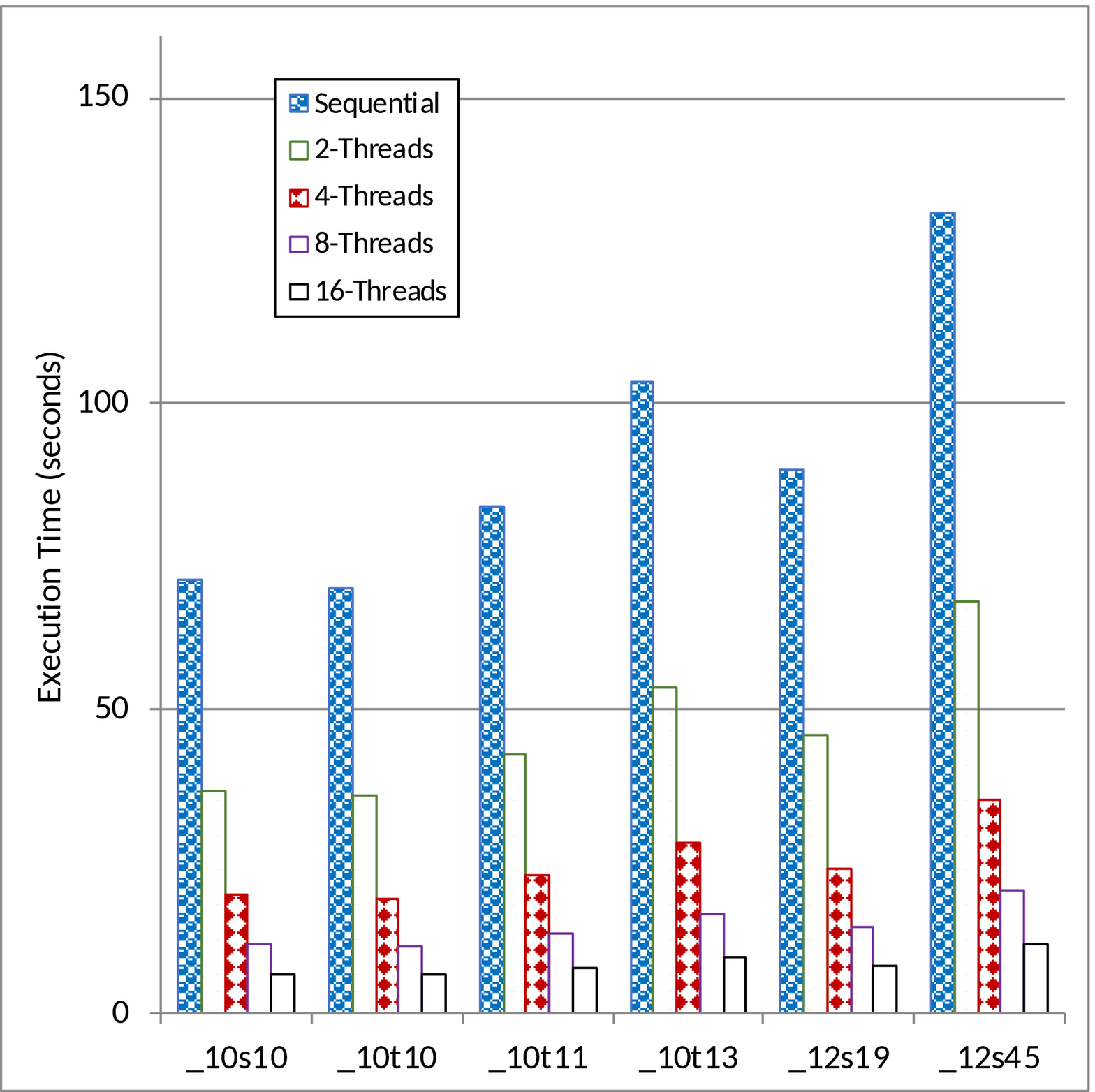}}

\caption{Parallel MSAProbs total execution time with different number of threads.}
\label{fig:execution-time-masprobs}
\end{figure}

The {\em MSAProbs} algorithm also shows a high scalability with a different number
of threads on the given set of files. Figure~\ref{fig:execution-time-masprobs}
shows the performance on both Balibase data files, i.e., 	Figure~\ref{fig:msaprobs-balibase},
and OXbench data files, i.e., Figure~\ref{fig:msaprobs-oxbench}. 
		
Figure~\ref{fig:msaprobs-balibase} shows that {\em MSAProbs} is able to achieve around $9X$ speedup using 16 threads
compared to sequential implementation across different files from the Balibase dataset. Further,
Figure~\ref{fig:msaprobs-oxbench} shows an average speedup of $11X$ across the OXbench dataset's files.

\subsubsection{M2Align}

Figure~\ref{fig:m2align-balibase} shows the parallel execution performance of {\em M2Align} algorithm on Balibase dataset. 
The speedup from sequential execution to 16 threads executions are $7.85X$, $9.41X$, $7.50X$,
 and $10.55X$ in $BB20004$, $BB30013$, $BB40002$, $BB50012$ files, respectively.
  As shown,the parallel {\em M2Align} satisfies a higher speedup with larger number of threads
   compared to both {\em T-Coffee}, {\em MAFFT}, and {\em MSAprobs} algorithms. 

\begin{figure}[!ht]
  \centering
  \subfigure[Balibase dataset with parallel M2Align.]{
  \label{fig:m2align-balibase}
    \includegraphics[width=0.45\linewidth]{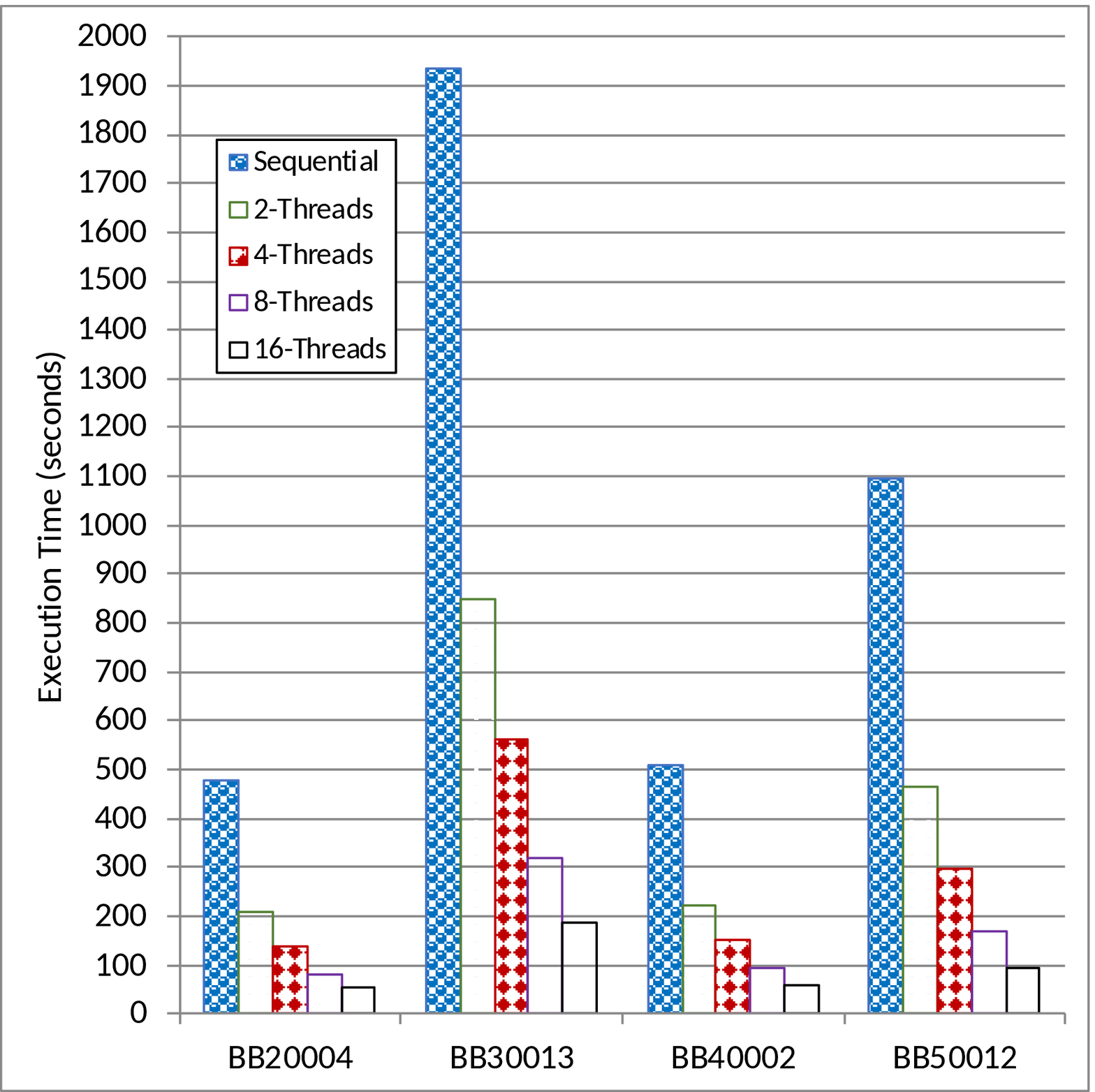}
    }
  \hfill
  \subfigure[OXbench dataset with parallel M2Align.]{
    \label{fig:m2align-oxbench}
   \includegraphics[width=0.45\linewidth]{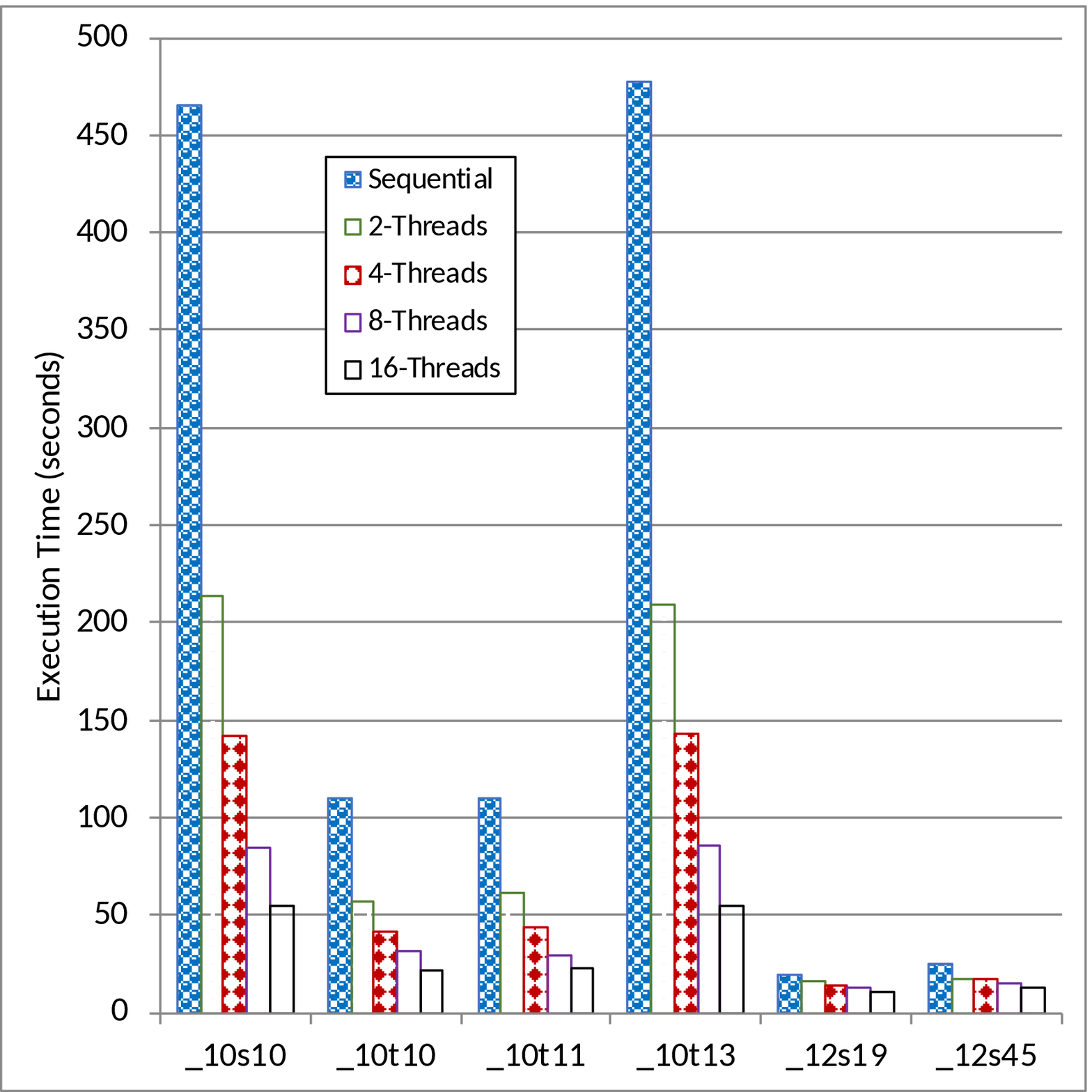}}

\caption{M2Align total execution time with different number of threads.}
\label{fig:m2align-balibase}
\end{figure}

The same case can also be shown by Figure~\ref{fig:m2align-oxbench}, where MSAProbs 
is used on OXbench dataset. The speedup is $7.43X$, $4.22X$m $3.91X$, $7.79X$, $0.82X$,
 and $0.91$ in $\_10s10$, $\_10t10$, $\_10t11$, $\_10t13$, $\_12s19$, $\_12s45$ files, respectively.

An important observation can be shown in Figure~\ref{fig:m2align-oxbench}.  With less sequential
 execution time, the scalability becomes worse and the speedup decreases, i.e., $\_12s19$, $\_12s45$ files.
	
\subsection{Accuracy Comparison}
Evaluating the efficiency of an MSA algorithm is related to the final alignment score 
it able to obtain. As mentioned before,  in this study we use three different scoring metrics, $Q/TC$, $cline$, $modeler$. All the scoring metrics use $Bench$ bench mark to estimate the accuracy score of different algorithms.

Table~\ref{tab:QTC_acc} shows the $Q/TC$ accuracy using our four target algorithms on nine different files from BAliBASE and OXBench datasets.  The average scoring values show that {\em T-Coffee} algorithm is the most efficient algorithm. {\em MSAProbs} algorithm comes in the second order followed by the {\em MAFFT} algorithm and finally the {\em M2align} algorithm.

\begin{table*}
\centering
\footnotesize

\caption{Accuracy comparison using $Q/TC$ score.}
\renewcommand{\arraystretch}{1.0}
\begin{tabu}{|c|[1pt]c|c|c|c|}
\hline
& \multicolumn{4}{c|}{Q/TC}  \\
Dataset & MAFFT	& T-Coffee & MSAProbs	  &	M2align \\ \hline

BB20004 &	$0.973/0.787$	& $0.984/0.787$	&$0.985/0.792$	 &$0.981/0.733$ \\ \hline
BB40002 &	$0.539/0$	&$0.99/0.737$ &	$0.908/0$ &	$0.574/0$ \\ \hline
BB50012 &	$0.866/0.429$	&$0.896/0.49$ &	$0.89/0.51$ &	$0.877/0.327$ \\ \hline
\_10s10 &	$0.926/0.667$&	$0.926/0.667$ &	$0.926/0.667$ &	$0.926/0.667$ \\ \hline
\_10t10 &	$0.833/0.741$	&$0.87/0.815$ &	$0.87/0.815$ &	$0.735/0.593$ \\ \hline
\_10t11	 &$0.603/0.429$	&$0.702/0.619$	 & $0.603/0.429$ &	$0.502/0.238$ \\ \hline
\_10t13&	$0.917/0.667$ &	$0.917/0.667$ &	$0.917/0.667$ &	$0.733/0$ \\ \hline
\_12s19	&$1$ &	$1$	&$1$	&$1$ \\ \hline
\_12s45&	$1$	&$1$&	$1$&	$1$ \\ \hline
Average &	$0.8507/0.6355$ &	$0.9205/0.7535$ &	$0.8998/0.6533$ &	$0.8142/0.506$\\ \hline
\end{tabu}
\label{tab:QTC_acc}
\end{table*}

Table~\ref{tab:cline_acc} shows the average accuracy scores of each algorithm using the same set of files. The tables shows that MSAProbs outperforms all other algorithms and T-Coffee algorithm is coming second followed by the MAFFT and the M2align algorithms.

\begin{table*}
\centering
\footnotesize

\caption{Accuracy comparison using {\texttt Cline} score.}
\renewcommand{\arraystretch}{1.0}
\begin{tabu}{|c|[1pt]c|c|c|c|}
\hline
& \multicolumn{4}{c|}{Cline}  \\
Dataset & MAFFT	& T-Coffee & MSAProbs	  &	M2align \\ \hline
BB20004	&	$0.572$	&	$0.576$	&	$0.578$	&	$0.576$	\\ \hline
BB40002	&	$0.1$	&	$0.25$	&	$0.317$	&	$0.113$	\\ \hline
BB50012	&	$0.38$	&	$0.399$	&	$0.406$	&	$0.378$	\\ \hline
\_10s10	&	$0.11$	&	$0.11$	&	$0.11$	&	$0.109$	\\ \hline
\_10t10	&	$0.281$	&	$0.291$	&	$0.292$	&	$0.262$	\\ \hline
\_10t11	&	$0.178$	&	$0.201$	&	$0.185$	&	$0.164$	\\ \hline
\_10t13	&	$0.0386$	&	$0.0386$	&	$0.0387$	&	$0.0332$	\\ \hline
\_12s19	&	$0.257$	&	$0.257$	&	$0.257$	&	$0.257$	\\ \hline
\_12s45	&	$0.434$	&	$0.434$	&	$0.436$	&	$0.434$	\\ \hline
Average	&	$0.2611$&	$0.2840$&	$0.2910$	&	$0.258$	\\ \hline

\end{tabu}
\label{tab:cline_acc}
\end{table*}

Table~\ref{tab:modeler_acc} also shows the average $modeler$ score of the algorithms over the same set of files.  The scores show that the {\em MSAProbs} algorithm satisfies the highest average score through all files. The {\em T-Coffee} algorithm satisfies the second highest score followed by the {\em MAFFT} and the {\em M2align}  algorithms.

\begin{table*}
\centering
\footnotesize

\caption{Accuracy comparison using {\texttt Modeler} score.}
\renewcommand{\arraystretch}{1.0}
\begin{tabu}{|c|[1pt]c|c|c|c|}
\hline
& \multicolumn{4}{c|}{Modeler}  \\
Dataset & MAFFT	& T-Coffee & MSAProbs	  &	M2align \\ \hline	
BB20004	&	$0.4$	&	$0.402$	&	$0.404$	&	$0.402$	\\ \hline
BB40002	&	$0.0388$	&	$0.0752$	&	$0.149$	&	$0.0325$	\\ \hline
BB50012	&	$0.219$	&	$0.23$	&	$0.241$	&	$0.206$	\\ \hline
\_10s10	&	$0.0562$	&	$0.0564$	&	$0.0563$	&	$0.0582$	\\ \hline
\_10t10	&	$0.161$	&	$0.168$	&	$0.168$	&	$0.15	$\\ \hline
\_10t11	&	$0.09$	&	$0.104$	&	$0.0908$	&	$0.0763$	\\ \hline
\_10t13	&	$0.0193$	&	$0.0193$	&	$0.0194$	&	$0.0158$	\\ \hline
\_12s19	&	$0.147$	&	$0.147$	&	$0.147$	&	$0.147$	\\ \hline
\_12s45	&	$0.278$	&	$0.278$	&	$0.278$	&	$0.278$	\\ \hline
Average	&	$0.1565$	&	$0.1644$	&	$0.1726$	&	$0.1517$	\\ \hline

\end{tabu}
\label{tab:modeler_acc}
\end{table*}

\subsection{Speedup Comparison}
Increasing the number of workers in parallel execution usually leads
to improve the performance of the underlying application. In this subsection,
we evaluate the gained speedup of increasing the number of cores.
Because {\em M2Align} algorithm gives the lowest execution time through
all given data files, we use it as a benchmark of the performance of the other
algorithms. The main goal of this experiment is to show if a particular algorithm
speedup increases with a larger number of cores.

 \begin{figure}
\centering
\includegraphics[width=0.7\linewidth]{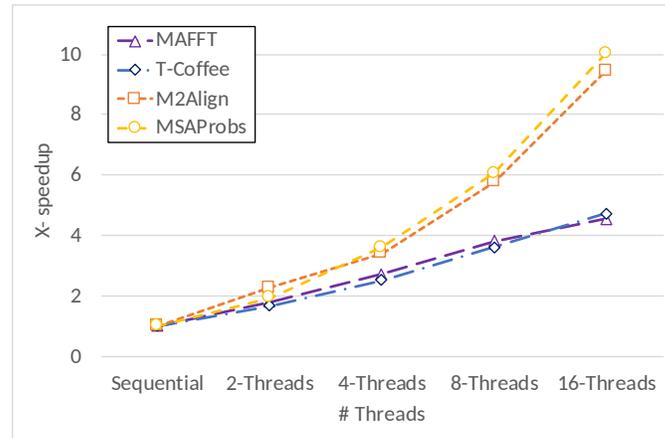}
  \caption{{\em T-Coffee}, {\em MAFFT}, {\em MSAProbs}, and {\em M2Align} average
   speedup comparison using different number of threads on different data files.}
  \label{fig:speedup-fig}
\end{figure}

Figure~\ref{fig:speedup-fig} illustrates the average gained
 speedup using different number of threads compared to the
  sequential version of each algorithm on different data files. 
  
As shown, both {\em MSAProbs} and {\em M2Align} algorithms can respectively gain upto
          1.93X and 2.27X speedup using 2-threads,
          3.61X and 3.41X speedup using 4-threads,
          6.07X and 5.79X speedup using 8-threads,
 and 10.04X and 9.46X speedup using 16-threads. Both algorithms show super-linear scalability  with larger number of threads.

Further, {\em T-Coffee} and {\em MAFFT} algorithms can respectively gain upto
           1.67X and 1.80X speedup using 2-threads,
          2.52X and 2.73X speedup using 4-threads,
          3.62X and 3.82X speedup using 8-threads,
 and 4.72X and 4.56X speedup using 16-threads.  Both algorithms show linear scalability  with larger number of threads.

 The figures shows the strong scalability of both {\em MSAProbs} and {\em M2Align} compared to both {\em T-Coffee} and {\em MAFFT} algorithms.

\section{Conclusion}
\label{sec:con}
The literature reveals a lack of providing a detailed comparison between existing parallel MSA algorithms
to show their strengths and weaknesses. Thus, in this work, our main objective is to study the state-of-the-art parallel MSA algorithms
from two different perspectives, performance and accuracy.
We picked up four different algorithms {\em T-Coffee}, {\em MAFFT}, {\em MSAProbs},
and {\em M2Align} and gave a detailed explanation of each algorithm and the available
 parallel implementation on multicore systems.

We also  gave a detailed evaluation of selected algorithms. The evaluation shows linear 
scalability of the four algorithms up to 16 cores with two different dataset files, Balibase and OXbench.




\end{document}